\documentclass[a4paper]{jpconf}
\usepackage{graphicx}
\usepackage{placeins}

\usepackage{hyperref}

\newcommand{\alicept}{\ensuremath{p_{\rm T}}}
\newcommand{\aliceptjet}{\ensuremath{p_{\rm{T,ch.jet}}}}

\newcommand{\pp}{pp}
\newcommand{\PbPb}{\ensuremath{\mbox{Pb--Pb}}}
\newcommand{\pPb}{\ensuremath{\mbox{p--Pb}}}
\newcommand{\Raa}{\ensuremath{R_{\rm{AA}}}}
\newcommand{\Iaa}{\ensuremath{I_{\rm{AA}}}}

\begin{document}

\title{Overview of ALICE results on azimuthal correlations using neutral- and heavy-flavor triggers}

\author{Sona Pochybova for the ALICE Collaboration}

\address{MTA Wigner RCP, $29-33$ Konkoly-Th\' ege Mikl\' os street, $1121$ Budapest, Hungary}

\ead{sona.pochybova@cern.ch}

\begin{abstract}
The ALICE detector is dedicated to studying the properties of hot and dense matter created in heavy-ion collisions. Among the probes used to investigate these properties are high-momentum particles, which originate in hard-scatterings occurring before the fireball creation. The fragments of hard scatterings interact with the hot and dense matter and via this interaction their spectra and azimuthal distributions are modified. This is probed by the measurement of the nuclear modification factor, where the \alicept\ spectra obtained in \PbPb\ collisions are compared to a \pp\ baseline. A strong suppression of charged hadrons as well as neutral- and heavy-flavor mesons was observed at $\alicept > 4$ GeV/$c$. Azimuthal correlations, using high-momentum ($\alicept > 4$ GeV/$c$) hadrons as triggers, can provide further insight into how the presence of the medium modifies the final kinematic distributions of the particles. Comparison with theoretical models can be used to test their predictions about the properties of the medium.
We give an overview of ALICE azimuthal-correlation measurements of neutral- and heavy-flavor mesons with charged hadrons in \pp\ collisions at $\sqrt{s}=7$ TeV and \PbPb\ collisions at $\sqrt{s_{\rm{NN}}}=2.76$ TeV. We also present a measurement of the $\pi^{0}$ correlation with jets in \pp\ collisions at $\sqrt{s}=7$ TeV. 

\end{abstract}

\section{Introduction}
In the high-energy heavy-ion collisions the creation of a new state of matter is expected. This matter is the hot and dense "primordial soup" called the \textit{Quark-Gluon Plasma (QGP)} \cite{Bazavov:2011nk}. The ALICE (A Large Ion Collider Experiment) collaboration at the LHC is dedicated to the study of the properties of this new state of matter  \cite{Alessandro:2006yt}. To do this, a set of probes are exploited. Among these are the high-momentum particles that were created from the fragments of high-momentum partons produced in the early stages of the collision. These partons may travel through the fireball probing its evolution. Partons lose energy via medium-induced gluon radiation \cite{Gyulassy:1990ye,Baier:1996sk} or via elastic collisions with the medium constituents \cite{Thoma:1990fm,Braaten:1991jj}. Once the fireball freezes out, these partons hadronize into high-momentum particles that carry the information about the parton-medium interaction. We can look at how the properties of these high-momentum particles change with respect to simpler systems as \pp\ and \pPb\ collisions and obtain information about the properties of the matter they emerged from and test our expectations.

In this work, we focus on the measurements using high-momentum $\pi^{0}$ and $D$ mesons as trigger particles. High-momentum neutral pions are experimentally identified via the $2$-photon decay channel in a wide range of momentum and can be used to scan different regions. The $D$ mesons consist of charm quarks, that are created in the initial stages of the collision. Due to their large masses, they are predominantly produced in hard scatterings. The energy loss these heavy quarks suffer, is expected to differ from light quarks and gluons. The gluon radiation at small angles with respect to the quark direction is suppressed in heavy-quark fragmentation, which produces a \emph{"dead-cone"} around the original parton. This implies a smaller energy loss for heavier quarks. The dead-cone effect is also present in vacuum \cite{Dokshitzer:2001zm}.

One of the observables sensitive to the parton-medium interaction is the \textit{Nuclear modification factor}:

\begin{equation}
\label{eqRAA}
\Raa(\alicept) = \frac{1}{\langle T_{\rm{AA}}\rangle} \frac{\rm{d}N_{\rm{AA}}/\rm{d}\alicept}{\rm{d}\sigma_{\mathrm{pp}}/\rm{d}\alicept},
\end{equation}

where $\langle T_{\rm{AA}}\rangle$ is the nuclear overlap function evaluated in the Glauber model \cite{Miller:2007ri}, proportional to the number of binary nucleon-nucleon collisions $\langle N_{\rm{coll}}\rangle$. By comparing the yields extracted in heavy-ion and \pp\ collisions, this variable tests whether the heavy-ion collision is a mere superposition of independent nucleon-nucleon collisions. If this would be the case, the $\Raa(\alicept)$ should be 1. Previous measurements showed a yield-suppression at high momenta for unidentified hadrons \cite{Aamodt:2010jd} as well as $D$ mesons \cite{ALICE:2012ab} and $\pi^{0}$ \cite{Abelev:2014ypa}. This suppression is generally understood as a manifestation of the energy loss the initial partons suffer when they traverse the QGP.

Another observable, that complements the modification of the momentum spectrum is the per-trigger yield obtained from angular correlations:

\begin{equation}
\label{eqCorr}
C(\Delta\eta\Delta\varphi)=\frac{1}{N_{\rm{trig}}}\frac{\rm{d} N_{\rm{assoc}}}{\rm{d}\Delta\eta\rm{d}\Delta\varphi}.
\end{equation}

This analysis considers hadrons with $\alicept\ > 8$ GeV/$c$ as high-momentum triggers. We take the trigger-hadron and investigate the angular distribution of the associated hadrons in different momentum bins. In this way, we can observe how the associated particle distributions are modified by the presence of the medium. When a high-virtuality process occurs it produces, at leading order, two back-to-back partons. If this process occurs close to the surface of the fireball, one of the partons will have to travel a shorter distance through the hot and dense matter than the other. Therefore the correlation analysis suffers a "\textit{surface-bias}" that allows us to study path-length dependence of the parton energy-loss. The high-momentum particles associated with the parton close to the surface are identified as triggers and particles associated with this parton's energy-loss appear on the near-side ($\Delta\varphi \approx 0$). The other parton has to travel through the fireball experiencing large energy-loss and producing associated particles on the away-side ($\Delta\varphi \approx \pi$).
To quantify the modification of the particle-yields in a certain momentum interval associated with the trigger, we construct the \textit{per-trigger yield modification factor (\Iaa)}:

\begin{equation}
\label{eqIaa}
	\Iaa(\Delta\varphi, \alicept) = \frac{Yield(\PbPb)}{Yield(\mathrm{pp})}.
\end{equation}

Apart from the yield-modification, one can also look at the changes in the shape of the angular distribution of the associated particles produced along the path of the trigger. Such measurement targets the broadening and softening of the initial parton fragmentation and is done by measuring the widths of the near- and away-side peaks.

In the following, we present the results of the analysis on $\pi^{0}$ angular correlation with charged particles $(h^{\pm})$ in \pp\ and \PbPb\ collisions at $\sqrt{s}=2.76$ TeV per-nucleon and $\pi^{0}$-jet angular correlations in \pp\ collisions at $\sqrt{s}=7$ TeV. The correlations of $D$ mesons and charged particles in \pp\ collisions at $\sqrt{s}=7$ TeV are compared with results from \pPb\ collisions at $\sqrt{s_{NN}}=5.02$ TeV and PYTHIA Monte Carlo simulations \cite{Sjostrand:2006za,Sjostrand:2014zea,Skands:2010ak}.

\section{Analysis setup}

The ALICE detector is a powerful system of sub-detectors, that was designed to investigate the properties of QGP in a broad range of variables and kinematics. A detailed description can be found in Ref.~\cite{Aamodt:2008zz}. In the presented analysis, the following sub-detectors were used: for the tracking and identification of the charged particles the ITS \cite{Aamodt:2010aa}, TPC \cite{Alme:2010ke} and TOF \cite{Akindinov:2010zza} were used covering the central pseudo-rapidity region $|\eta|<0.9$ and full azimuth $\varphi=2\pi$. The $\pi^{0}$s were reconstructed using the EMCal \cite{Bourrion:2012vn} electromagnetic calorimeter with the acceptance $|\eta| < 0.7$ and $\Delta\varphi=100^{\circ}$. In the $\pi^{0}$-jet analysis, jets were reconstructed using the $\rm{anti-}k_{\rm{T}}$ jet-finding algorithm \cite{Cacciari:2011ma}. For the reconstruction only charged tracks have been used in $|\eta|<0.9$ and jets with $R=0.4$ and $|\eta|<0.5$ were accepted.

\subsection{Identification of $\pi^{0}$}

The neutral pions were identified using the shower-shape topological analysis in the EMCal. For energies larger than $5-6$ GeV, the electromagnetic showers produced in the calorimeter by the two photons from $\pi^{0}$ decay merge into a single large cluster of calorimeter cells (a cell is the smallest unit  of the calorimeter, with $6\times 6$ cm$^{2}$ area at $4.28$ m from the interaction point). By studying the energy profile of the cluster, we can determine whether it was produced by a single photon or if it is a superposition of two. This procedure allows us to measure $\pi^{0}$s in the EMCal up to $50$ GeV.
 
\subsection{Identification of $D$ mesons}
For the $D$-meson triggered correlation analysis the $D^{+}, D^{0}$ and $D^{*+}$ have been reconstructed in three decay modes: $D^{0}\rightarrow K^{-}\pi^{+}$, $D^{+}\rightarrow K^{-}\pi^{+}\pi^{+}$ and $D^{*}(2010)^{+}\rightarrow D^{0}\pi^{0}$ \cite{Nakamura:2010zzi}. A detailed description of the procedure can be found in Ref.~\cite{ALICE:2011aa} and the references therein. The selection procedure exploits the displacement of the decay vertex with respect to the primary vertex. Kaons and pions were identified using TPC and TOF detectors and from these the $D$-meson invariant mass was reconstructed.

\subsection{Setup of the correlation studies}

In the correlation analysis, the triggers were associated with the unidentified hadrons and charged jets. For each trigger-associated pair, we calculate $\Delta\varphi=\varphi_{\rm{trig}}-\varphi_{\rm{assoc}}$ and $\Delta\eta=\eta_{\rm{trig}}-\eta_{\rm{assoc}}$ in the signal event. To correct for the detector $\Delta\varphi$ and $\Delta\eta$ acceptance, we divide the raw ($\Delta\varphi \Delta\eta$) correlation with the mixed-event correlation. This is a correlation, that is not physical and is constructed by taking a trigger from the signal event and correlating it with an associated particle from another, but topologically similar event. Dividing the signal correlation by the mixed, we obtain the pair-acceptance corrected distribution. This distribution consists of two peaks, which sit on top of a pedestal, that is not correlated with the trigger and has different physics origins. In these analyses, we want to only study the per-trigger yield in the peaks so we have to subtract this pedestal. In \pp\ collisions, the background is assumed to be flat and is determined using \textit{Zero Yield At Minimum (ZYAM)} assumption \cite{Ajitanand:2005jj}. In \PbPb\ collisions, the situation becomes more complicated, since the background is expected to be modulated by the initial azimuthal anisotropy called elliptic flow \cite{Adler:2002ct}. Therefore, the background is subtracted in the analysis of \PbPb\ data using both a flat and a flow assumption.

\section{Results and discussion}

In this section, we present our results on the angular correlation analyses using $\pi^{0}$ and $D$ mesons as triggers. First, the results of $\pi^{0}$-hadron and $\pi^{0}$-jet correlation are shown, then, we proceed with $D$-meson analyses. The plots include both systematic and statistical errors, which are not combined but are plotted separately.

\subsection{The $\pi^{0}$ triggered correlations}

The $\pi^{0}$ - charged particle correlation analysis was performed in \pp\ collisions and $10\%$ most central \PbPb\ collisions at a per-nucleon collision energy of $2.76$ TeV.

\begin{figure}[tb]
\begin{minipage}{0.47\textwidth}
\includegraphics[width=\textwidth]{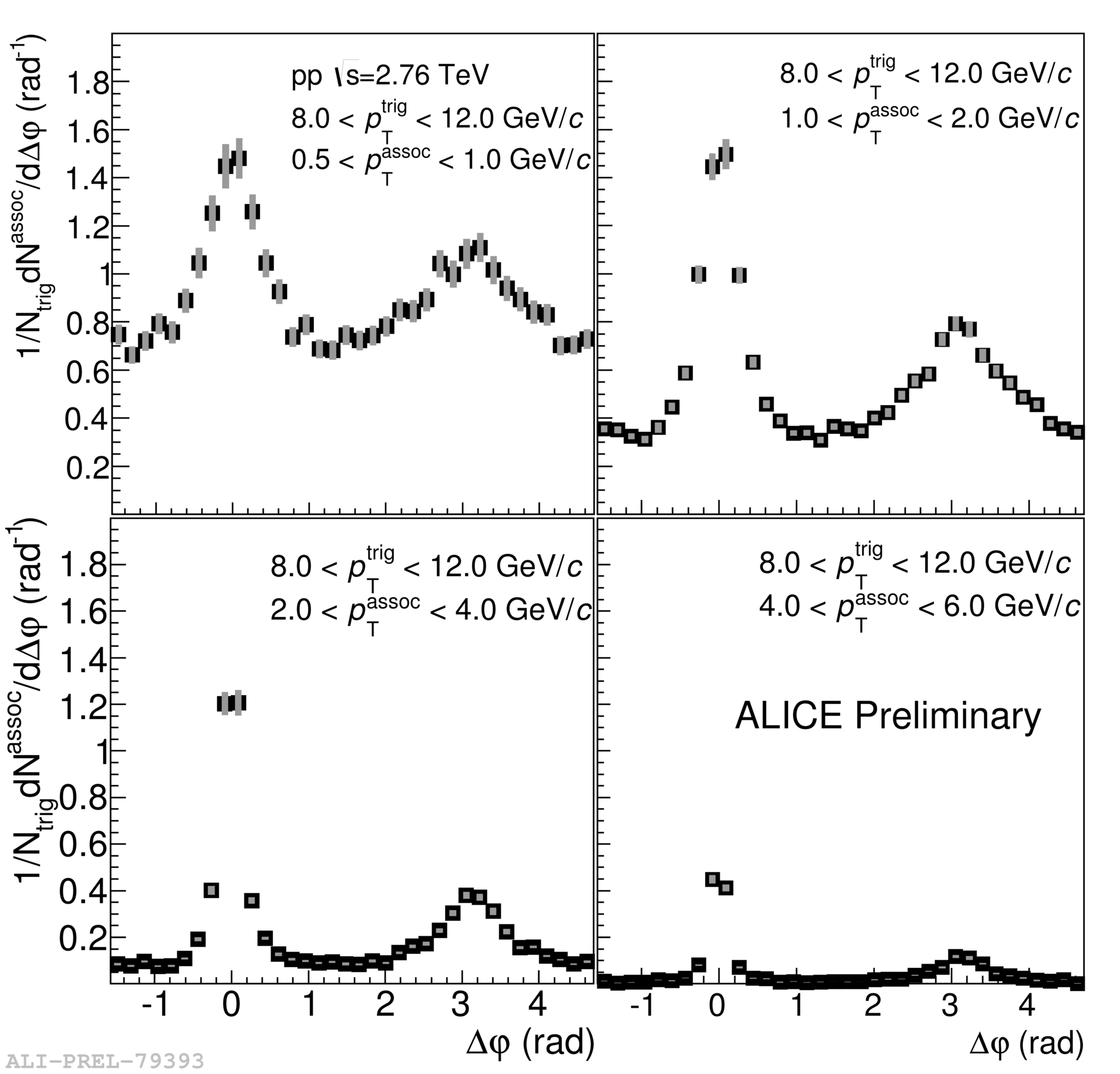}
\begin{center}
(a)
\end{center}
\end{minipage}
\hspace{0.06\textwidth}%
\begin{minipage}{0.47\textwidth}
\includegraphics[width=\textwidth]{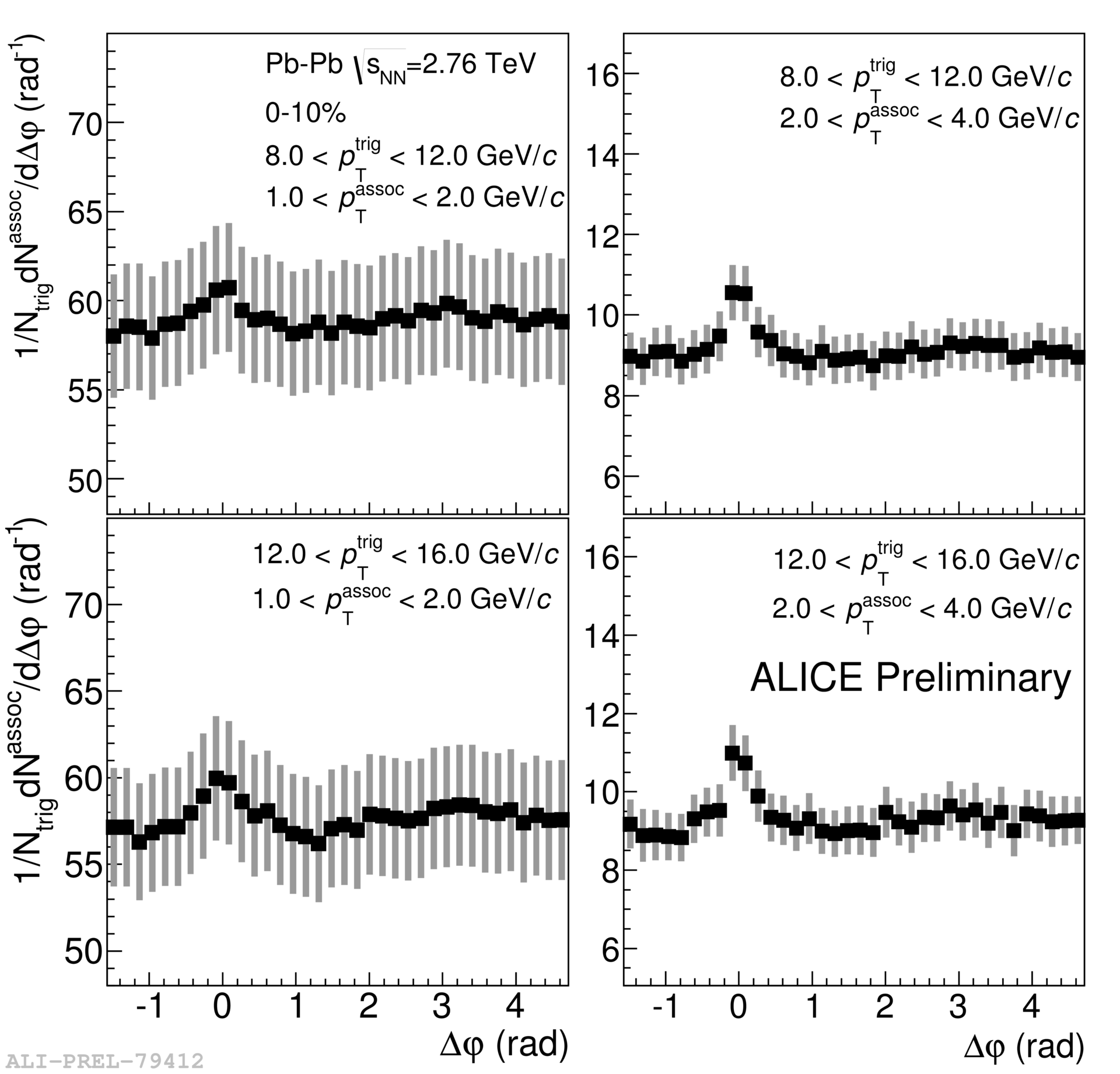}
\begin{center}
(b)
\end{center}
\end{minipage}
\caption{\label{pi0dPhippPbPb}The $\pi^{0}$ -charged particle correlation: angular distribution of the per-trigger yield for different colliding systems and different momentum of $\pi^{0}-h^{\pm}$ pairs without background subtraction. The distributions are corrected for the pair-acceptance. Panel (a) shows results from \pp\ collisions at $\sqrt{s}=2.76$ TeV with $8.0<\alicept^{\rm{trig}}<12.0$ GeV/$c$ and $\alicept^{\rm{assoc}}\in\{(0.5;1.0),(1.0;2.0),(2.0;4.0),(4.0;6.0) \}$ GeV/$c$. Panel (b) shows the per-trigger yield in \PbPb\ collisions at $\sqrt{s_{\rm{NN}}}=2.76$ TeV. The momentum ranges of the associated pairs are $\alicept^{\rm{assoc}}\in\{(1.0;2.0),(2.0;4.0) \}$ GeV/$c$ and the ranges of trigger-momentum are $8.0<\alicept^{\rm{trig}}<12.0$ GeV/$c$ and $12.0<\alicept^{\rm{trig}}<16.0$ GeV/$c$ at top and bottom of the figure, respectively.}
\end{figure}

Figure \ref{pi0dPhippPbPb} shows the evolution of the angular distribution of the per-trigger yield as it changes with the momentum of the trigger-associate pair and with colliding system from \pp\ to \PbPb\ collisions. The distributions are corrected for the pair-acceptance but are not corrected for the background below the jet-peaks. We observe that with increasing momentum of the associated particles, this background becomes suppressed and the jet peaks start to become the dominant features in the distribution. This behavior is common to both \pp\ and \PbPb\ systems, but is more pronounced in the former.

\begin{figure}[tb]
\begin{minipage}{0.47\textwidth}
\includegraphics[width=\textwidth]{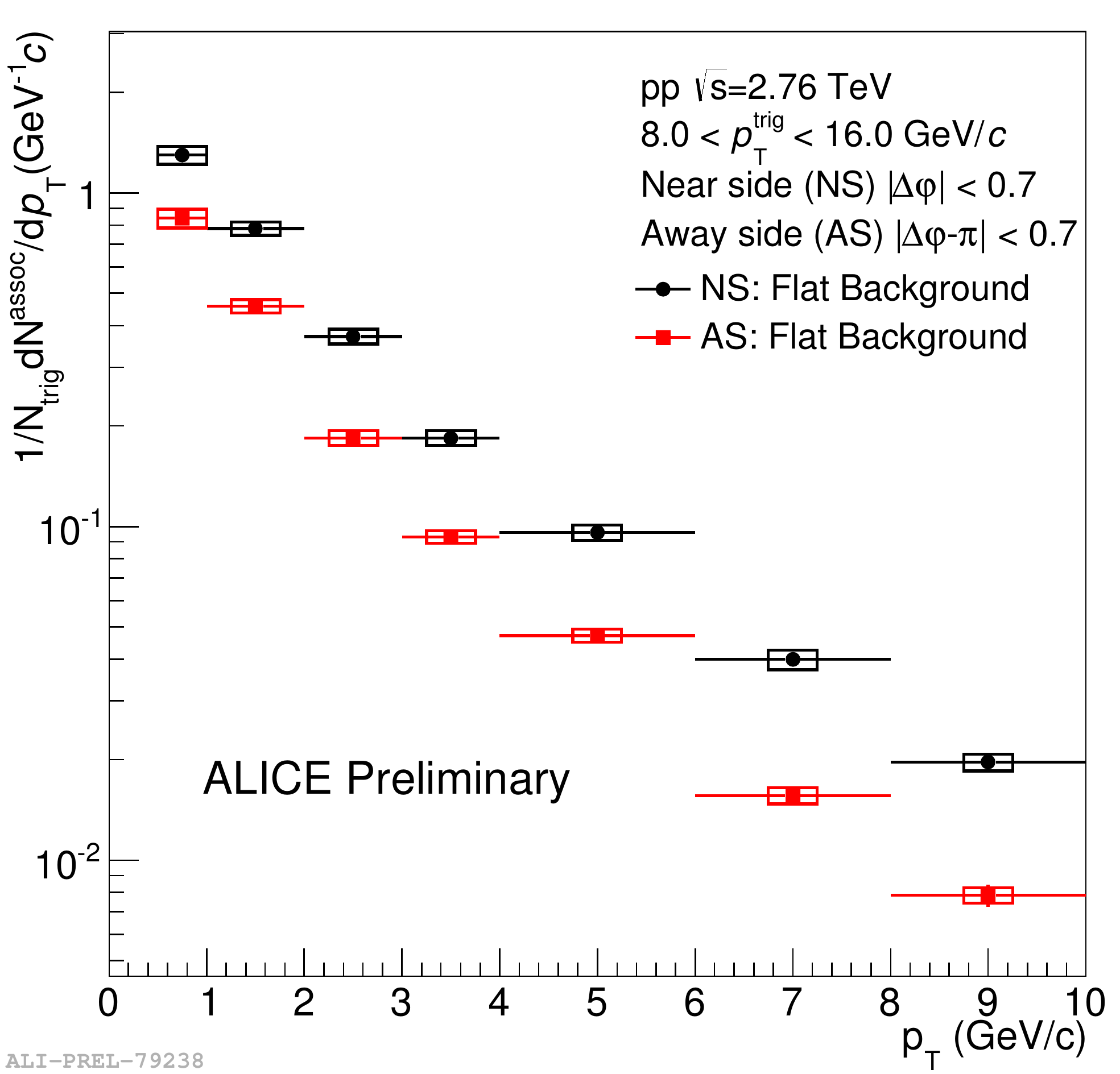}
\end{minipage}
\hspace{0.06\textwidth}%
\begin{minipage}{0.47\textwidth}
\includegraphics[width=\textwidth]{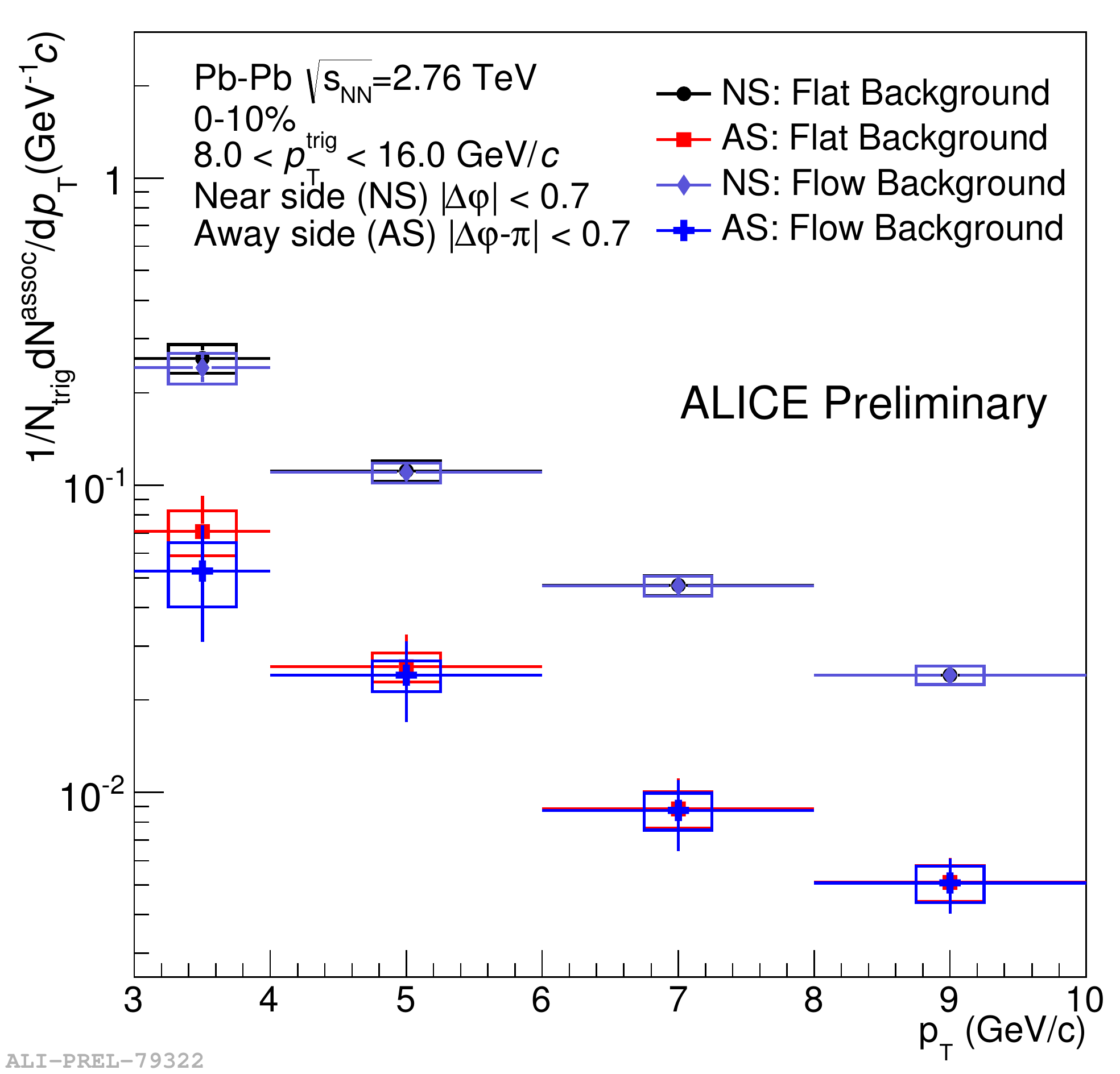}
\end{minipage}
\caption{\label{pi0YieldNSASppPbPb}The $\pi^{0}-h^{\pm}$ correlation: per-trigger yield on the near- and away-side as a function of associated track momentum in \pp\ collisions at $\sqrt{s}=2.76$ TeV (left) and in $10\%$ most central \PbPb\ collisions at $\sqrt{s_{\rm{NN}}}=2.76$ TeV (right). The background subtracted in \pp\ collisions is flat. In \PbPb\ collisions, the yields are compared for flat and flow background assumption.}
\end{figure}

Once background is subtracted and only the jet-peaks remain, the per-trigger yield is extracted in the $\alicept^{\rm{assoc}}$ bins for the near- and away-side. The size of the integration range in $\Delta\varphi$ is 0.7 and is the same on both sides. In the case of \pp\ collisions, only the flat background assumption was considered, whereas in the case of \PbPb\ collisions, a flow background assumption was added. The results are shown in Figure \ref{pi0YieldNSASppPbPb}. The yield decreases with momentum of the associated tracks and is smaller on the away-side. In the \PbPb\ case, the extracted yields are consistent for both background assumptions within errors.

\begin{figure}[bt]
\begin{minipage}{0.47\textwidth}
\includegraphics[width=\textwidth]{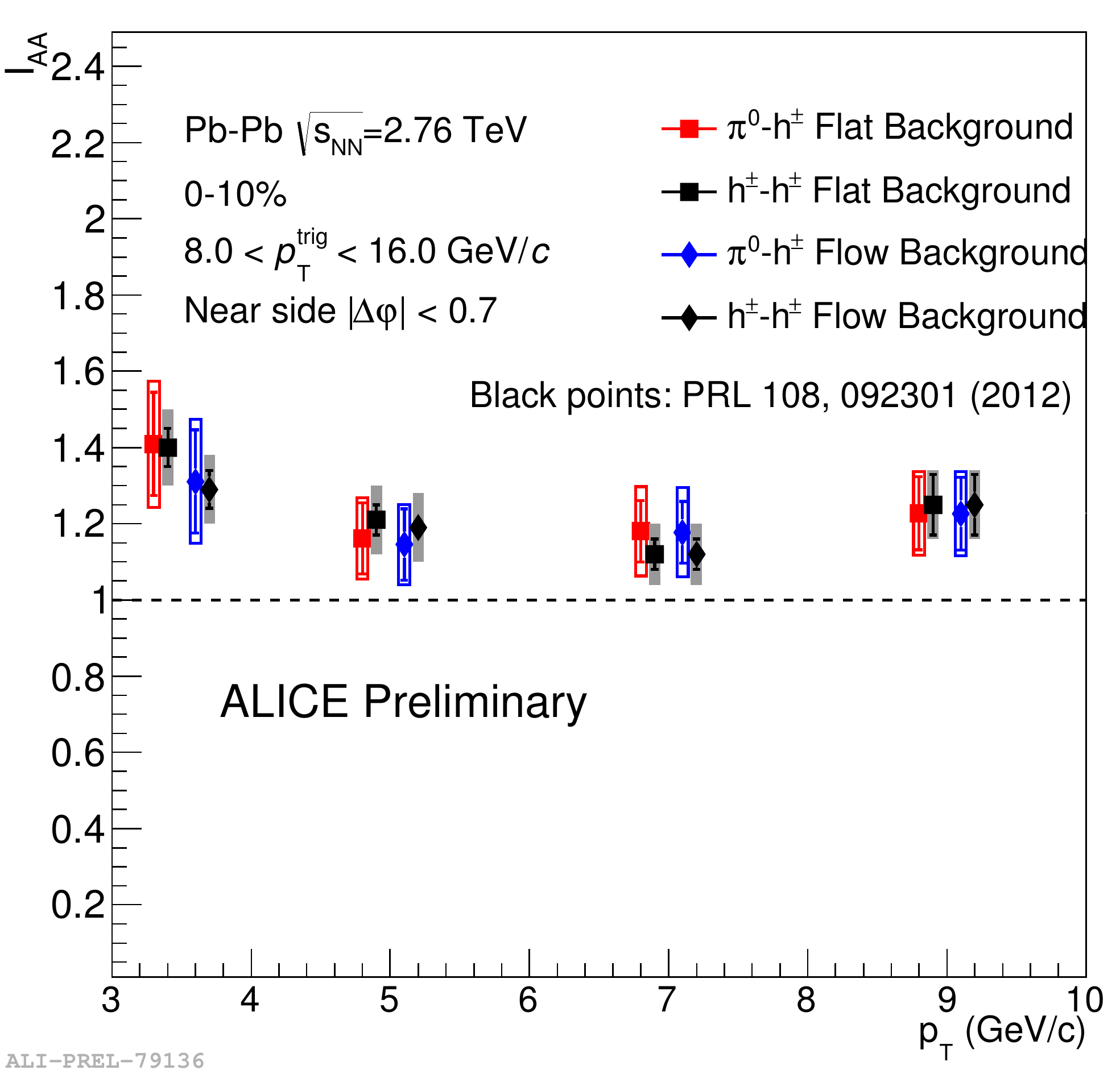}
\end{minipage}
\hspace{0.06\textwidth}%
\begin{minipage}{0.47\textwidth}
\includegraphics[width=\textwidth]{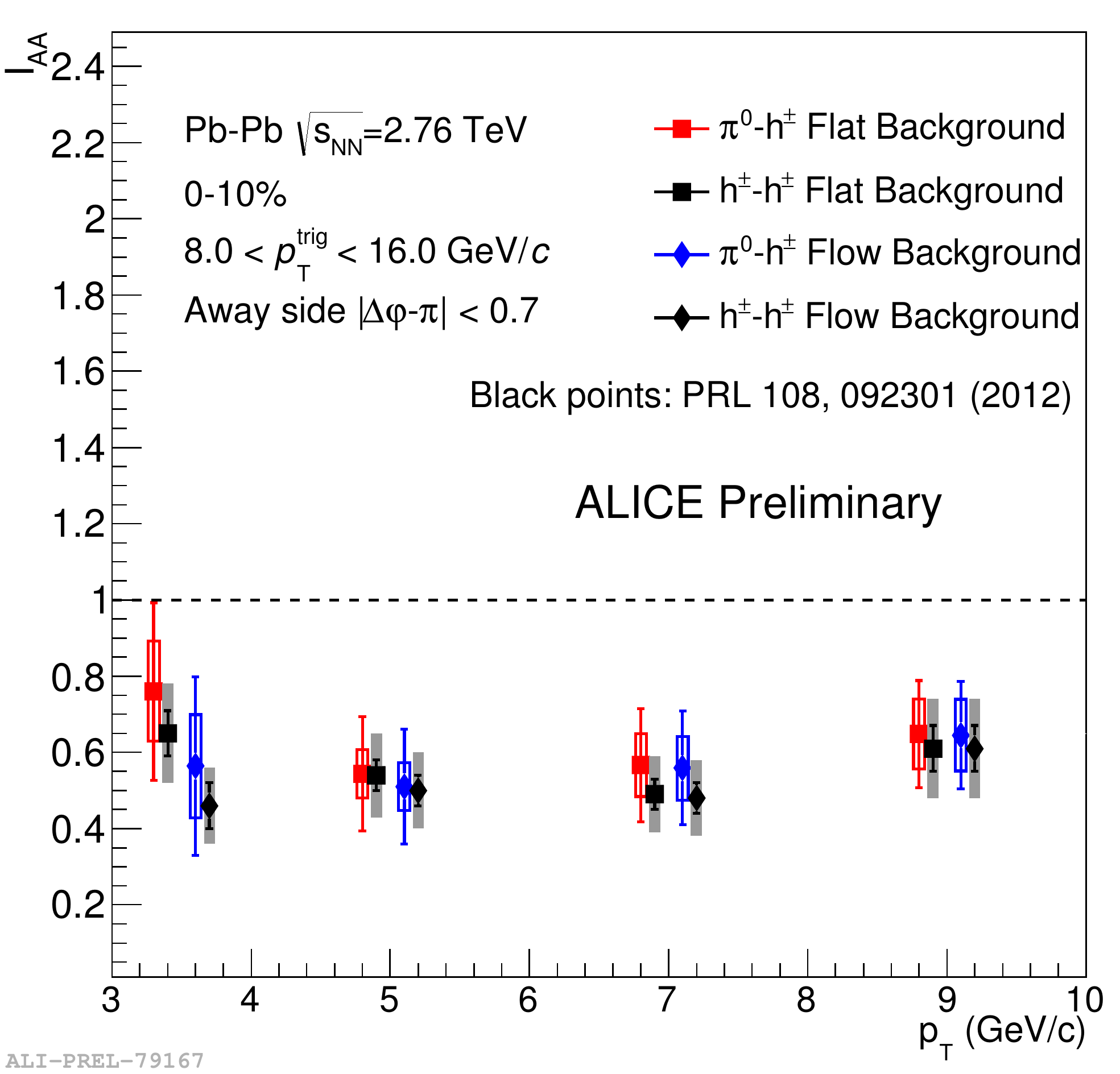}
\end{minipage}
\caption{\label{pi0IAANSAS}Per-trigger yield modification factor (\Iaa) as a function of the \alicept\ of the associated particle on the near-side (left) and away-side (right) for $10\%$ most central \PbPb\ collisions at $\sqrt{s_{\rm{NN}}}=2.76$ TeV. The results of the $\pi^{0}-h^{\pm}$ analysis are compared to the results of $h^{\pm}-h^{\pm}$ analysis for a flat and a flow background assumption.}
\end{figure}

To quantify the difference between the behavior of the $\pi^{0}-h^{\pm}$ correlation in \pp\ and \PbPb\ collisions, the \Iaa\ (defined by eq.~\ref{eqIaa}) is calculated from the extracted yields. The $\alicept^{\rm{assoc}}$ dependence of this variable is shown in Figure \ref{pi0IAANSAS}. On the near-side, we observe a $\approx 20\%$ enhancement while on the away-side a $\approx 50\%$ suppression. The \Iaa\ is independent of momentum and background assumption within errors. The results are compared to the results of $h^{\pm}-h^{\pm}$ analysis \cite{Aamodt:2011vg} and are found to be consistent within errors.

The suppression on the away-side can be understood in terms of medium-induced parton energy-loss mentioned in the introduction. It is interesting to point out, that this suppression is smaller than the one measured by the STAR collaboration at RHIC \cite{Adams:2006yt} ($\approx 75\%$). The difference in the suppression might indicate a difference in the biases of the correlation analysis between the two experiments. A discussion on how various biases of the experimental correlation measurements influence the results and their interpretation can be found in \cite{Renk:2012ve}. Based on this, the main contributors to the differences may be the interplay between the following effects: \textit{(i)} different "surface bias" of the measurement \textit{(ii)} larger momentum transfer in hard-scattering processes at LHC and \textit{(iii)} change of the fraction of quarks and gluons produced in the final state.

Due to larger collision energies, the partons created in the hard-scatterings have a higher initial energy and it is likely for them to survive the fireball with shifted momentum even if they have to travel through the whole length of the medium. Apart from this, even if the hard-scattering occurs further away from the surface, the partons might be energetic enough to have the particles produced in their fragmentation triggered in the experimental setup. Additionally, the fraction of quarks and gluons created in the final state of hard-scatterings changed when going from RHIC to LHC energies. While the quarks dominated in the RHIC regime, it is more likely to produce gluons at the LHC. Gluons have a softer fragmentation, thus one can expect them to be more prone to interact with the surrounding medium and as a result produce a large number of particles. The higher initial parton energy combined with a softer fragmentation and a higher probability to interact with medium may result in an apparently smaller suppression in a given momentum range. 

Looking at the near-side, the STAR collaboration measured an \Iaa\ consistent with unity. The consistency with unity suggests, that very little interaction between the trigger and the surrounding matter occurred. The presented ALICE measurement showed a clear enhancement, indicating that the trigger has been affected by its travel through the medium. Following a similar line of thought to the previous paragraph, one can understand this via the difference between the kinematics of the hard-scattering processes between the RHIC and the LHC. Apart from identifying triggers at the surface, one identifies also the ones originating inside the fireball. These suffer the medium-induced energy-loss and are identified as triggers in a lower momentum bin together with an abundance of particles around them, that were produced as the result of the parton-medium interaction. The fact supporting that the difference in the \Iaa\ between STAR and ALICE has the same source on the near- and away-side is that ALICE adds $\approx 20\%$ to both sides of the STAR result. 

Other possible interpretation of this enhancement is the medium-induced softening of the fragmentation function, discussed briefly in Ref.~\cite{Aamodt:2011vg}. Hadrons in a given momentum range, which were produced in such modified fragmentation, originate from higher-momentum partons in \PbPb\ collisions as compared to \pp.\ This difference between the two systems is then translated into $\Iaa > 1$.

\begin{figure}[tb]
\includegraphics[width=\textwidth]{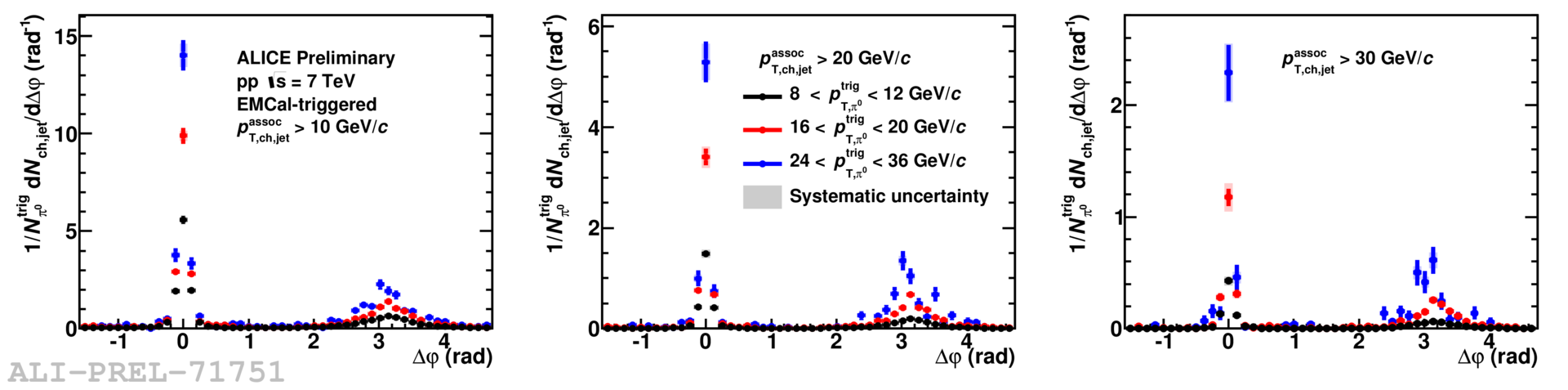}
\caption{\label{pi0JetCorr}Angular distribution of per-trigger jet-yield from the $\pi^{0}$-jet correlation analysis in \pp\ collisions at $\sqrt{s}=7$ TeV. The distributions are compared for three trigger-momentum bins: $\aliceptjet^{\rm{trig}}\in\{ (8;12), (16;20), (24;36) \}$ GeV/$c$. For the associated jet three momentum thresholds are defined: $\aliceptjet^{\rm{assoc}}>10$ GeV/$c$ (left), $\aliceptjet^{\rm{assoc}}>20$ GeV/$c$ (middle) and $\aliceptjet^{\rm{assoc}}>30$ GeV/$c$ (right).}
\end{figure}

To further test these ideas, one would need to conduct an analysis targeting directly the jet-fragmentation via the measurements of hadron-jet correlations. Such measurements have already started. The $\pi^{0}$-jet correlation analysis was performed in \pp\ collisions at $\sqrt{s}=7$ TeV. Three trigger-momentum bins and three jet-momentum thresholds were studied. The angular distributions of the per-trigger jet-yield are shown in Figure~\ref{pi0JetCorr}. We observe clear peaks on both the near- and away-side. The jet-yield increases with the momentum of $\pi^{0}$ and decrease with the jet-momentum threshold.  

\begin{figure}[tb]
\includegraphics[width=\textwidth]{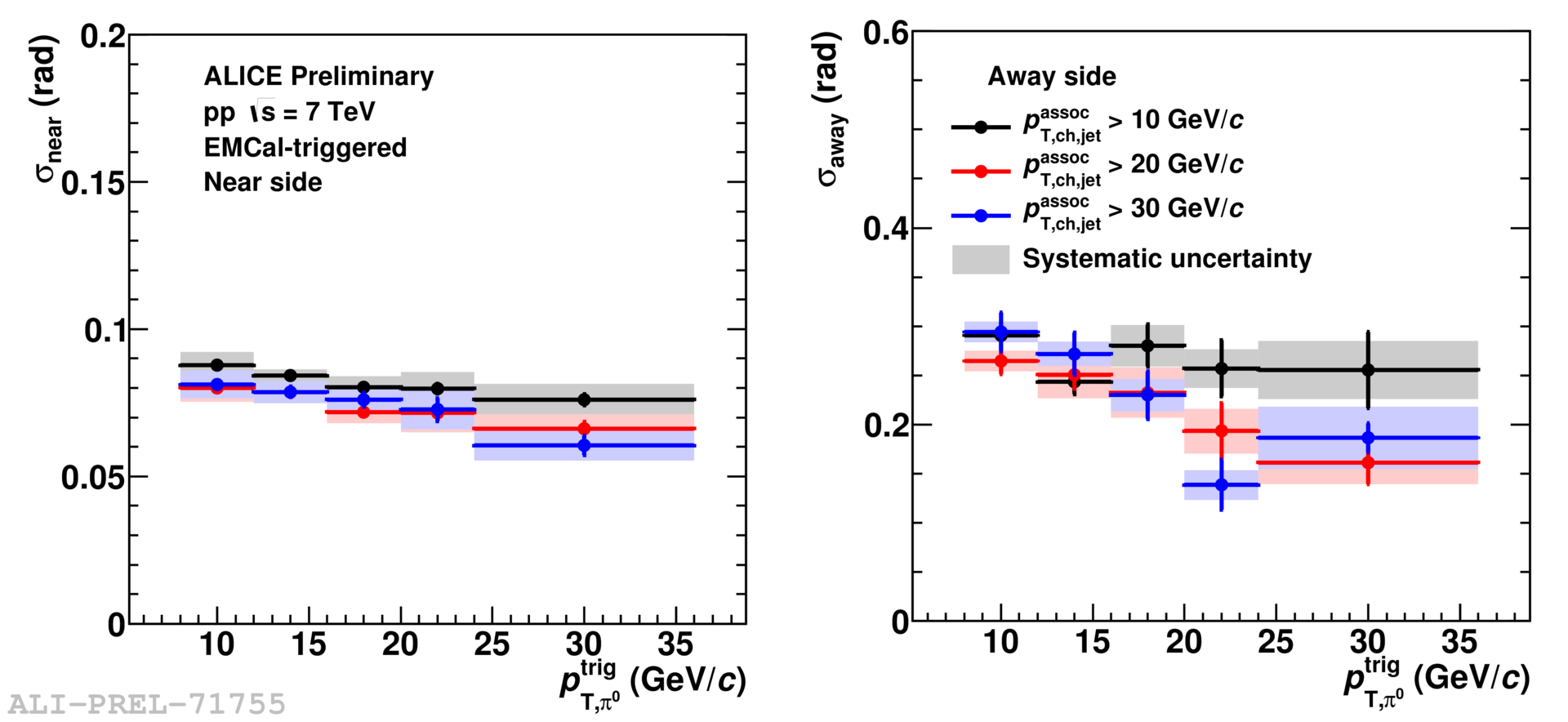}
\caption{\label{pi0JetCorrWidth}Widths of the angular distribution of per-trigger jet-yield from the $\pi^{0}$-jet analysis in \pp\ collisions at $\sqrt{s}=7$ TeV. The widths are plotted as a function of $\pi^{0}$ momentum for three associated jet-momentum thresholds: $\aliceptjet^{\rm{assoc}}>10, 20, 30$ GeV/$c$. The widths are extracted on the near- (left) and away-side (right).}
\end{figure}

Figure~\ref{pi0JetCorrWidth} shows the widths of the peaks and their dependence on $\alicept^{\rm{trig}}$ and $\aliceptjet^{\rm{assoc}}$. Within errors, the widths are independent of associated jet momentum-threshold and they exhibit a decreasing trend with the trigger-particle momentum.

These results show that the meson is produced closer to the jet-axis as its momentum increases. On the other hand, the width of the angular distribution seems to be independent of jet's threshold momentum. This indicates a small jet-momentum dependence of jet-fragmentation into $\pi^{0}$s. However, it is important to mention that this measurement was done using only charged jets, and so it is difficult to establish a direct relationship between the charged-jet and neutral-pion momentum. A more robust claim on the jet's fragmentation into $\pi^{0}$s requires a full-jet reconstruction with both the charged and the neutral part.

\subsection{$D$ meson - charged particle azimuthal correlations}

\begin{figure}[tb]
\begin{minipage}{0.47\textwidth}
\includegraphics[width=\textwidth]{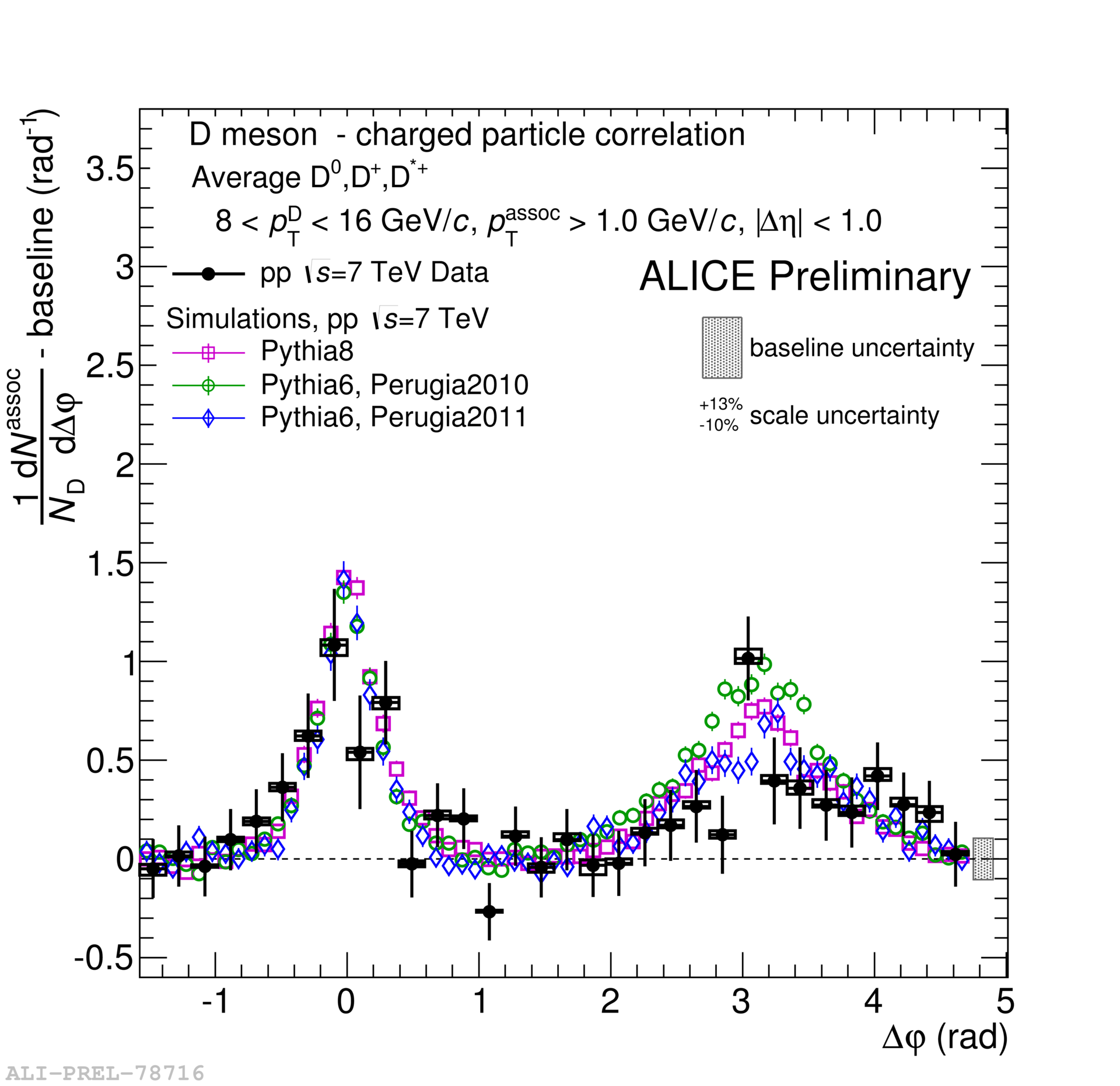}
\end{minipage}\hspace{0.06\textwidth}%
\begin{minipage}{0.47\textwidth}
\includegraphics[width=\textwidth]{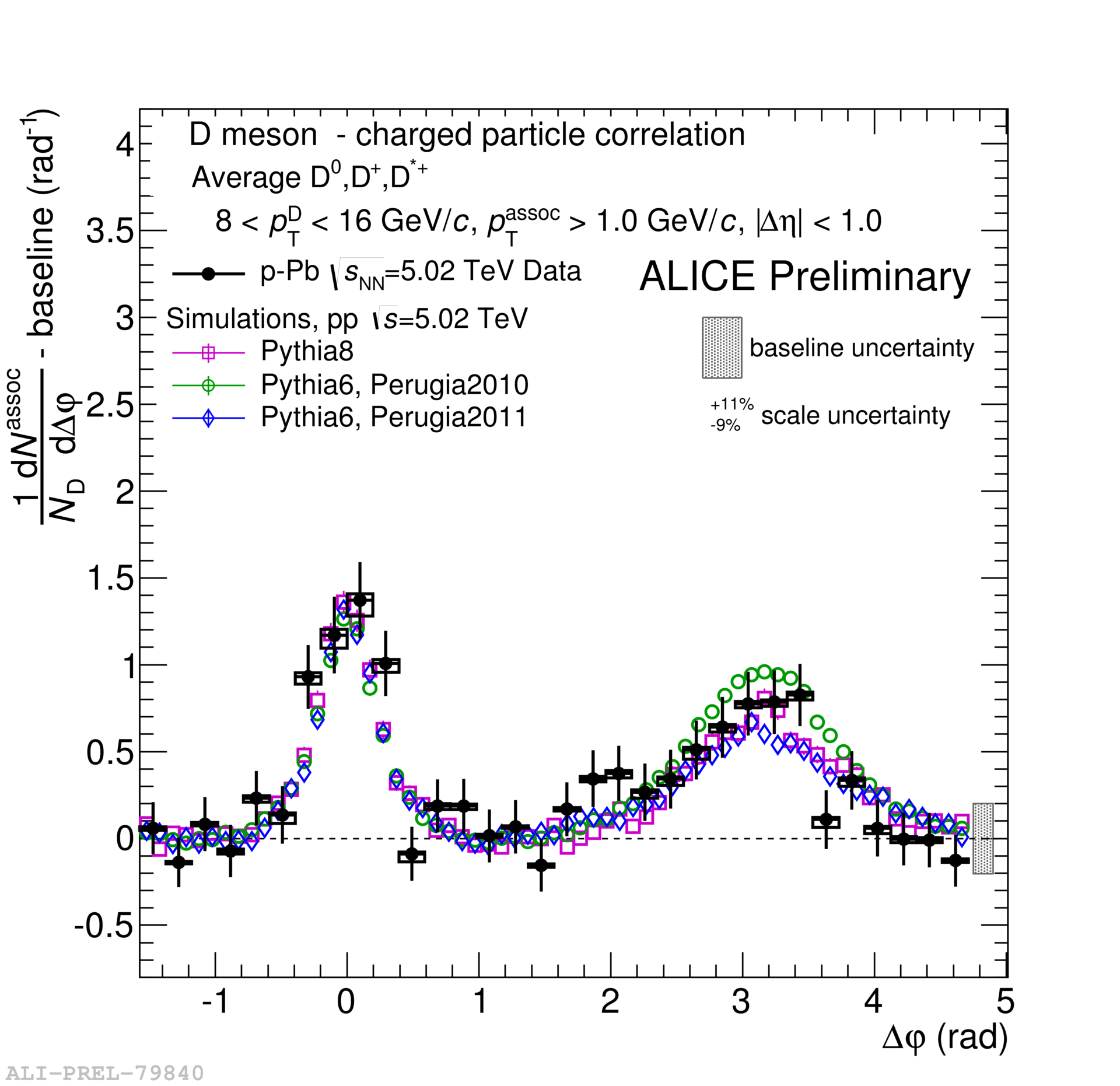}
\end{minipage}
\caption{\label{hDpppPbMC}The $D$ meson - charged particle azimuthal correlations:: angular distribution of per-trigger yield in \pp\ collisions at $\sqrt{s}=7$ TeV (left) and \pPb\ collisions at $\sqrt{s_{\rm{NN}}}=5.02$ TeV (right). The trigger used is a $D$ meson with transverse momentum $8.0<\alicept^{\rm{D}}<16$ GeV/$c$. The associated tracks are unidentified particles with $\alicept^{\rm{assoc}}>1.0$ GeV/$c$. The distributions are compared to Monte Carlo simulations of \pp\ collisions at $\sqrt{s}=7$ TeV in the case of \pp\ collisions, and $\sqrt{s}=5.02$ TeV in the case of \pPb\ collisions. The PYTHIA 8 and PYTHIA 6 generators are used, the latter implementing two tunes: Perugia 2010 and Perugia2011.}
\end{figure}

Azimuthal correlations of $D$ mesons and charged particles were studied in \pp\ collisions at $\sqrt{s}=7$ TeV and \pPb\ collisions at $\sqrt{s_{NN}}=5.02$ TeV. The $D$ mesons were identified in the range $8<\alicept^{D}<16$ GeV/$c$ and were correlated to unidentified charged particles with $\alicept^{\rm{assoc}}>1$ GeV/$c$. The per-trigger yield as a function of $\Delta\varphi$ is shown in Figure \ref{hDpppPbMC}. The distributions are compared to PYTHIA Monte Carlo simulations of \pp\ collisions at corresponding center-of-mass energies: PYTHIA 8 \cite{Sjostrand:2014zea} and PYTHIA 6 with Perugia 2010 and Perugia 2011 tunes \cite{Skands:2010ak}. Both \pp\ and \pPb\ results are within errors reproduced by the Monte Carlo models.

\begin{figure}[tb]
\begin{minipage}{0.47\textwidth}
\includegraphics[width=\textwidth]{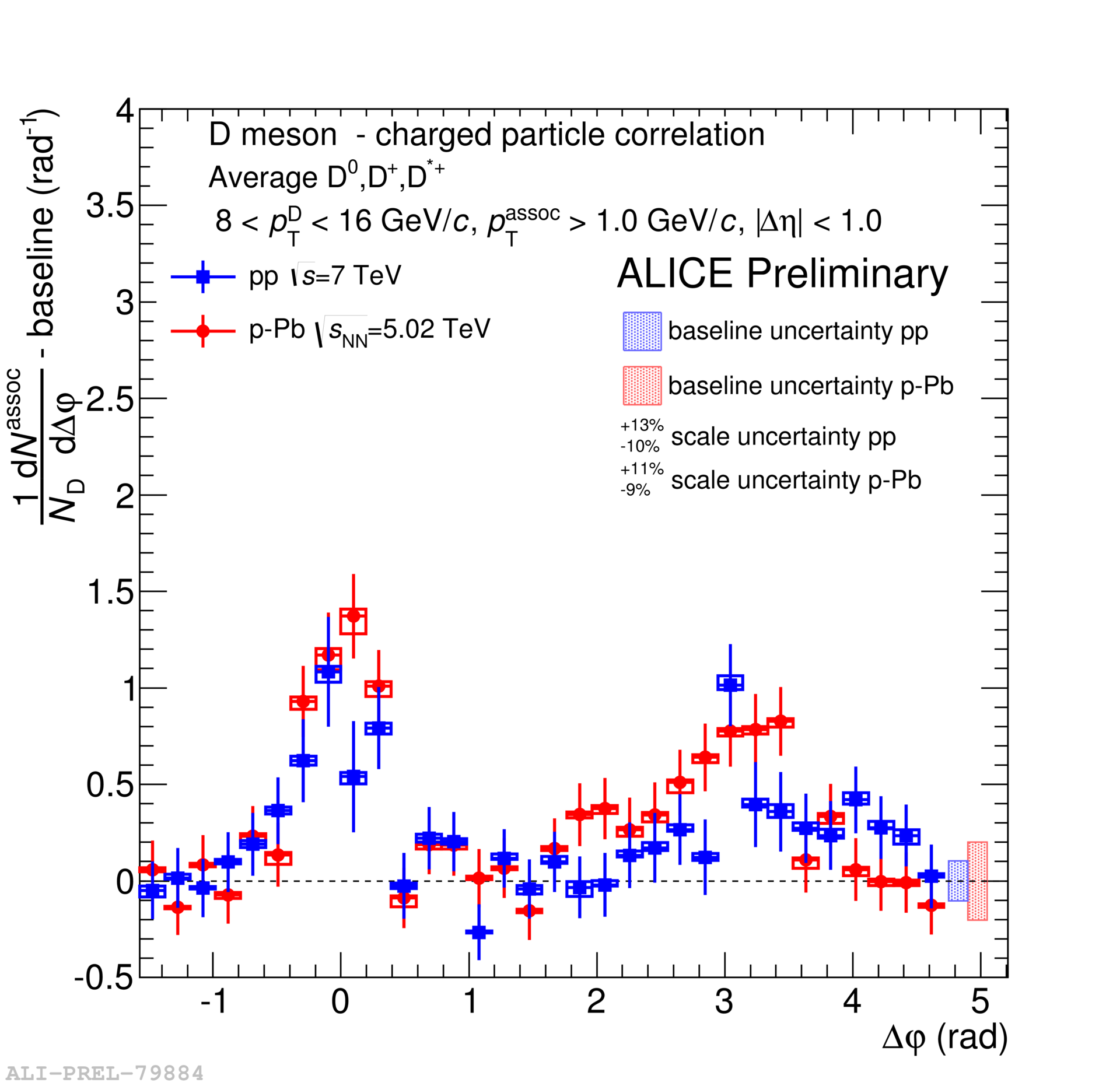}
\end{minipage}\hspace{0.06\textwidth}%
\begin{minipage}{0.47\textwidth}
\includegraphics[width=\textwidth]{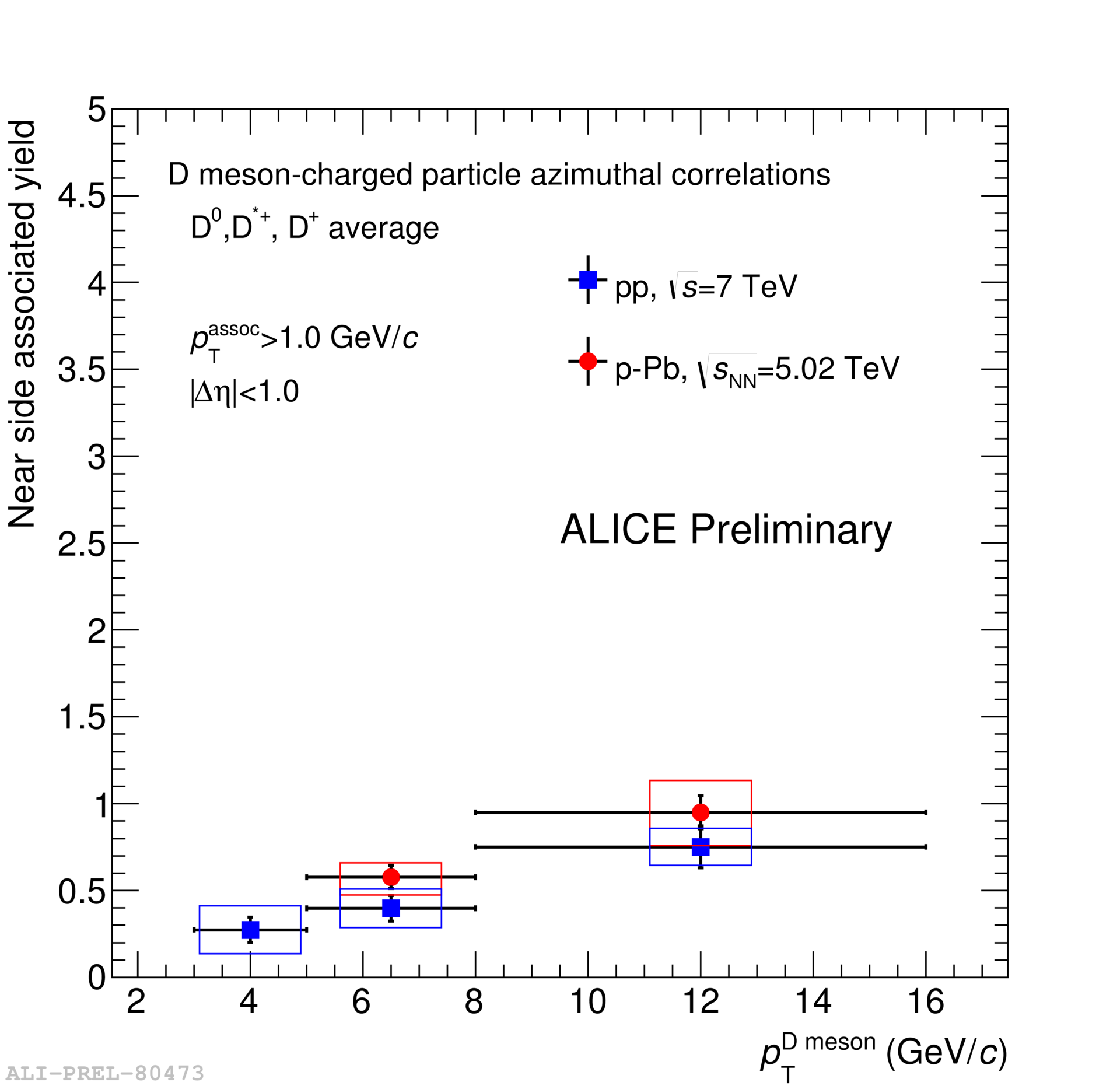}
\end{minipage}
\caption{\label{hDppPb}Comparison of $D$ meson - charged particle correlations in \pp\ collisions at $\sqrt{s}=7$ TeV and \pPb\ collisions at $\sqrt{s}=5.02$ TeV.  Left: Comparison of the angular distribution of per-trigger yield.
Right: Comparison of near-side per-trigger yield as a function $D$-meson momentum.}
\end{figure}

Figure~\ref{hDppPb} shows a direct comparison of angular distribution of per-trigger yield (left) and near-side yield as a function of $\alicept^{D}$ (right) in pp and \pPb\ collisions. The results obtained in the two collision systems agree within errors. However, due to large uncertainties it is difficult to draw strong conclusions about the comparison.

\section{Conclusions and final remarks}

We have presented an overview of the results on the correlation-measurements with $\pi^{0}$ and $D$ mesons as high-\alicept\ triggers. We can conclude, that the results from $\pi^{0}$-triggered correlations in \PbPb\ collisions are consistent with the presence of hot and dense matter that causes partons to lose energy as they travel through it. This interaction is affected by the kinematics of the partons created in the initial stages of the collision as can be seen from the comparison of RHIC and LHC results. A ore detailed investigation of the jet-fragmentation modification has already begun.

In the $D$-meson correlation measurements, first results are available in pp and \pPb\ collisions, setting an important baseline for upcoming heavy-ion analyses. In \pPb\ collisions, no indication of hot nuclear effects were observed. In the upcoming Run II, the statistical uncertainties are expected to significantly improve. The ITS upgrade planned for Run III should provide an improved spatial resolution for the secondary vertex recognition, which is crucial for the performance in heavy-flavor measurements.

\subsection{Acknowledgments}

I would like to thank the ALICE Collaboration for giving me the opportunity to present the results. Further, I would like to thank the Wigner RCP and the OTKA grant NK 116109 for providing necessary funding for my attendance of the $10^{\rm{th}}$ International Workshop on High-pT Physics in the RHIC/LHC era in Nantes, France ($9^{\rm{th}}-12^{\rm{th}}$ September $2014$).

\section*{References}
\bibliographystyle{unsrt}
\bibliography{references}
 
\end{document}